
\input PHYZZX
\titlepage
\sequentialequations

\title{QCD Phase Transition at Finite Temperature
in the Dual Ginzburg-Landau Theory}
\author{
H.~Ichie$^{\rm a}$
\foot{e-mail
address : ichie@atlas.phys.metro-u.ac.jp},
H.~Suganuma$^{\rm b}$ and H.~Toki$^{\rm b,c}$
}
\address{
{\rm a)} Department of Physics, Tokyo Metropolitan University,
\break
Hachiohji, Tokyo 192, Japan
}
\address{
{\rm b)} Institute for Physical and Chemical Research (RIKEN),
\break
Wako, Saitama 351-01, Japan
}
\address{
{\rm c)} Research Center for Nuclear Physics (RCNP),
Osaka University,
\break
Ibaraki, Osaka 567, Japan
}

\abstract{
We study the pure-gauge QCD phase transition
at finite temperatures in the dual
Ginzburg-Landau theory, an effective theory of
QCD based on the dual Higgs mechanism.
We formulate the effective potential
at various temperatures by introducing the quadratic source term,
which is a new useful method to obtain the effective potential
in the negative-curvature region.
Thermal effects reduce the QCD-monopole condensate and
bring a first-order deconfinement phase transition.
We find a large reduction of the self-interaction
among QCD-monopoles and the glueball masses
near the critical temperature
by considering the temperature dependence of the self-interaction.
We also calculate the string tension at finite temperatures.
}

\endpage


\REF\cheng{
For instance,
T.~P.~Cheng and L.~F.~Li,
``Gauge theory of elementary particle physics",
(Clarendon press, Oxford, 1984) 1.
}
\REF\rothe{
For instance, H.~J.~Rothe, ``Lattice Gauge Theories",
(World Scientific, Singapore, 1992) 1.
}
\REF\suzuki{ T.~Suzuki, Prog.~Theor.~Phys.~{\bf 80} (1988) 929 ;
 {\bf 81} (1989) 752.
 \nextline
 S.~Maedan and T.~Suzuki, Prog.~Theor.~Phys.~{\bf 81}
(1989) 229.
}
\REF\tHo{ G.~'t~Hooft, Nucl.~Phys.~{\bf B190} (1981) 455.
}
\REF\SuST{ H.~Suganuma S.~Sasaki and H.~Toki,
Nucl.~Phys.~{\bf B435} (1995) 207.
}
\REF\Kro{
A.~S.~Kronfeld, G.~Schierholz and ~U.~-J.~Wiese,
Nucl.~Phys.~{\bf B293} (1987) 461.
\nextline
T.~Suzuki and I.~Yotsuyanagi, Phys.~Rev. {\bf D42} (1990) 4257.
\nextline
S.~Hioki, S.~Kitahara, S.~Kiura, Y.~Matsubara,
O.~Miyamura, S.~Ohno and T.~Suzuki,
Phys.~Lett.~{\bf B272} (1991) 326.
}
\REF\latt{
For instance, H.~Shiba and T.~Suzuki,
Nucl.~Phys.~{\bf B} (Proc.~Suppl.) {\bf 34} (1994) 182.
}
\REF\TSS{
H.~Suganuma, S.~Sasaki and H.~Toki,
Proc. of Int. Conf. on
``Quark Confinement and Hadron Spectrum",
Como, Italy, (World Scientific, 1994).
\nextline
S.~Sasaki, H.~Suganuma and H.~Toki, {\it ibid}.
\nextline
H.~Toki, H.~Suganuma and S.~Sasaki,
Nucl.~Phys.~{\bf A577} (1994) 353c.
}
\REF\SaST{ S.~Sasaki, H.~Suganuma and H.~Toki,
RIKEN-AF-NP-172 (1994).
}
\REF\nielsen{
H.~B.~Nielsen and A.~Patkos, Nucl.~Phys.~{B195} (1982) 137.
}
\REF\kapusta{For a review article,
J.~I.~Kapusta, ``Finite-Temperature Field Theory",
Cambridge University Press, Cambridge, 1988) 1.
}
\REF\monden{ H.~Monden T.~Suzuki and Y.~Matsubara,
Phys.~lett.~{\bf B 294} (1992) 100.
}
\REF\ring{P.~Ring and P.~Schuck, ``The Nuclear Many-Body Problem'',
(Springer-Verlag, New York, 1980) 1.
}
\REF\karsch{ J.~Engels, F.~Karsch, H.~Satz and I.~Montvay,
Phys.~Lett.~{\bf B102} (1981) 332.
}
\REF\gao{ M.~Gao, Nucl.~Phys.~{\bf B9} (Proc. Suppl.) (1988) 368.
}


\FIG\EPa{ The effective potentials at various temperatures as
          functions of the QCD-monopole condensate $\bar \chi$.
          The numbers beside each curve are the temperatures.
          The absolute minimum points of the effective potentials are
          shown by crosses.}

\FIG\OPa{ The QCD monopole condensate $\bar \chi _{\rm phys}(T)$
 at minima of the effective potential
 as a function of the temperature $T$.
 The solid curve denotes $\bar \chi _{\rm phys}(T)$ corresponding
 to the confinement phase,
 which is the absolute minimum up to
 $T_C = $0.49 GeV and becomes a local minimum up to
 $T_{\rm up} = $0.51 GeV.
 A minimum appears at $\bar \chi=0$
 above $T_{\rm low} = $0.38 GeV and becomes the
 absolute minimum above $T_C = $0.49 GeV.
 The dot-dashed curve denotes
 the value of $\bar \chi$ at the local maximum.}

\FIG\OPb{
 The QCD monopole condensate $\bar \chi _{\rm phys}(T)$ at minima of the
 effective potential as a function of the temperature
 in the case of variable $\lambda (T)$
 so as to reproduce $T_C = $0.2GeV.
 The meanings of the curves are the same as in Fig.{\OPa}.
}

\FIG\GB{
The masses of the glueballs at various temperatures:
$m_B(T)$ and $m_\chi (T)$.
The solid  lines denote the case of variable $\lambda (T)$
with a constant $v$. The dashed lines denote the case of
variable $v(T)$ with a constant $\lambda $.
A large reduction of these masses is found near the critical
temperature. The dotted line denotes $m=T$.
The phase transition occurs at the temperature satisfying
$m_B, m_\chi  \simeq T$.
}

\FIG\STa{
The string tension $k(T)$ as a function of the temperature $T$.
The solid and dashed lines correspond to the variable $\lambda (T)$ case
with a constant $v$ and the variable $v(T)$ case with a constant $\lambda $,
respectively.
The constant ($\lambda $, $v$) case is also shown by the thin line.
The lattice QCD results in the pure gauge in Ref.[\gao]
are shown by black dots near the critical temperature.}


It is believed that the hadron physics is governed by the quantum
chromodynamics (QCD). Although the high energy phenomena as deep
inelastic scattering are described  by perturbative QCD owing to the
asymptotic freedom [\cheng], the theory becomes highly non-perturbative
at low energy. The lattice QCD theory [\rothe] was then developed
for low-energy phenomena and demonstrated again the goodness of QCD.
This gigantic numerical simulation method, however, does not tell us yet
how these low-energy phenomena as quark confinement and chiral symmetry
breaking take place. We need therefore some effective theory, which
incorporates the essence of the low-energy QCD physics and at the same
time reproduces the observables with a few parameters in the effective
theory, as the Ginzburg-Landau theory in the superconductivity.

In recent years, the Kanazawa group [\suzuki] proposed
the dual Ginzburg-Landau theory (DGL) as
an attractive effective theory of nonperturbative QCD.
The DGL theory incorporates the QCD monopole
fields as essential ingredients for confinement of colored particles
(quarks and gluons).
The QCD monopole field has its clear origin through
the abelian gauge fixing in the non-abelian gauge theory
a la 't~Hooft [\tHo].
In this theory, the QCD vacuum is characterized by QCD monopole
condensation, which provides a mass to the dual gauge field
through the dual Higgs mechanism
and hence the color electric field cannot freely develop in the
QCD vacuum.
Therefore, the color electric field originated from one colored
object should be confined in a small vortex-like tube to
end at the other colored object [\suzuki,\SuST].
This corresponds precisely to the Meissner effect of
superconductivity.
But, here the role of the magnetic and the electric fields
are reversed and
this phenomenon is called as the dual Meissner effect.
Such a picture for the color confinement has been
extensively investigated by recent studies [\Kro,\latt]
based on the lattice gauge theory,
and many evidences of QCD-monopole condensation
in the nonperturbative QCD vacuum have been reported [\Kro].

Suganuma, Sasaki and Toki (SST) studied the DGL theory
in further detail [\SuST].
They formulated the confinement potential of heavy quarks in a natural
way and obtained the linear potential in the same form as the
vortex-like solution in the superconductivity.
SST then showed QCD monopole condensation also
induced the dynamical chiral-symmetry breaking (D$\chi $SB)
[\SuST,\TSS,\SaST].
Thus, the DGL theory describes both the confinement and
D$\chi $SB of QCD in the non-perturbative region.
These two nonperturbative features, the confinement and D$\chi $SB, would
be changed in the high-temperature system, which is expected to be
realized as the quark-gluon-plasma (QGP)
in the ultra-relativistic heavy-ion
collisions or in the early universe [\kapusta].
Nowadays, the finite-temperature QCD including the QGP physics
is one of the most interesting subjects in the
intermediate-energy physics [\kapusta].

In this paper, we would like to develop
the thermodynamics of the DGL theory [\monden] and study
the change of the properties in the QCD vacuum
with temperatures especially in terms of the
deconfinement phase transition.
To this end, we concentrate on the pure-gauge QCD case, where
glueballs appear as the physical excitation.
Although such a pure gauge system is different from
the real world, it is regarded as a proto-type of the real QCD
and is well studied by using the lattice QCD simulation [\rothe].
It is worth mentioning that
our framework based on the DGL theory can be extended to include
the dynamical quarks straightforwardly [\suzuki,\SuST]
keeping the chiral symmetry of the system,
which is explicitly broken in the color-dielectric model [\nielsen]
or in the lattice QCD with the Wilson fermion [\rothe].

The DGL Lagrangian [\suzuki,\SuST,\monden]
relevant for the QCD vacuum is written
by the dual gauge field $\vec B_\mu =(B_\mu ^3,B_\mu ^8)$
and the QCD-monopole field $\chi _\alpha (\alpha =1,2,3)$,
$$
 {\cal L}_{\rm DGL}= - {1 \over 4}
(\partial_{\mu}\vec{B}_{\nu}-\partial_{\nu}\vec{B}_{\mu})^2 +
 \sum_{\alpha=1}^3[|(i\partial_{\mu}-g\vec{\epsilon_{\alpha}}
\cdot\vec{B}_{\mu})
\chi_{\alpha}|^2 - \lambda(|\chi_{\alpha}|^2-v^2)^2].
\eqn\La
$$
Here, $\vec \epsilon_\alpha $ denotes the magnetic charge of the
QCD-monopole field $\chi _\alpha $:
$\vec{\epsilon}_1 = (1,0),
\vec{\epsilon}_2 = (-{1 \over 2},-{\sqrt{3} \over 2})$ and
$\vec{\epsilon}_3 = (-{1 \over 2},{\sqrt{3} \over 2})$.
The dual gauge coupling constant $g$
satisfies the Dirac condition, $eg = 4\pi$,
with $e$ being the gauge coupling constant.
The strength $\lambda$ for the
self-interaction of the QCD-monopole field and
the vacuum expectation value $v$ are the parameters of the
DGL theory. In principle, the values of these parameters can
be extracted from the lattice QCD data,
but it is practically difficult at present.
Hence, these parameters are determined
by fitting to various low-energy observables.
The DGL Lagrangian {\La} is obtained by integrating over $\vec{A}_{\mu}$
in the original DGL partition functional in the pure gauge case
[\suzuki,\SuST].

We investigate the effective potential in the
path integral formalism.
The partition functional is written as
$$
  Z[J] = \int {\cal D}{\chi_{\alpha}}{\cal D}{\vec{B}_{\mu}}
\exp{\left( i\int d^4x \{{\cal{L}}_{\rm DGL}
-J\sum_{\alpha=1}^3|\chi_\alpha|^2\} \right) },
\eqn\Zj
$$
where we take the quadratic source term [\ring]
instead of the standard linear source term [\cheng,\kapusta].
As is well-known in the $\phi^4$ theory [\cheng,\kapusta],
the use of the linear source term
leads to an imaginary mass of the scalar field $\chi _\alpha $
in the negative-curvature region of the
classical potential, and therefore
the effective action cannot be obtained there due to
the appearance of ``tachyons".
In this respect, there is an extremely advanced point in
the use of the quadratic source term [\ring],
because the mass of the scalar field $\chi _\alpha $ is always real even
in the negative-curvature region of the classical potential
owing to the contribution of the source $J$
to the scalar mass. [See Eq.(6).]
Then, one obtains the effective action for the
whole region of the order parameter without any difficulty
of the imaginary-mass problem.
Moreover, the effective action with the quadratic source
can be formulated keeping the symmetry of the classical potential.
Since this method with the quadratic source term is quite general,
it is convenient to formulate the non-convex effective potential
in the $\phi ^4$ theory, the linear $\sigma $ model
or the Higgs sector in the unified theory [\cheng].

The vacuum expectation value of $\chi _\alpha $ ($\alpha $=1,2,3)
is the same value $\bar \chi $ due to the Weyl symmetry [\suzuki],
and therefore we separate the QCD-monopole field $\chi_\alpha $ into
its mean field $\bar \chi$ and
its fluctuation $\tilde{\chi}_\alpha $ as
$$
  \chi_\alpha = \left( \bar \chi + \tilde{\chi_\alpha} \right)
  \exp{\left( i\xi_\alpha \right)}.
\eqn\Chia
$$
Here, the phase variables $\xi_\alpha$ have a constraint,
$\sum_{\alpha =1}^3 \xi _\alpha =0$, where two independent degrees
of freedom remain corresponding to the dual gauge symmetry
$[{\rm U(1)}_3\times {\rm U(1)}_8]_m$ [\suzuki,\SuST].
When QCD monopoles condense, the phase variables $\xi_\alpha $
turn into the longitudinal degrees of
the dual gauge field $\vec B_\mu $, which is the dual Higgs mechanism.

Since we are interested in the translational-invariant system
as the QCD vacuum,
we consider the $x$-independent constant source $J$,
which leads to a homogeneous QCD-monopole condensate.
In the unitary gauge,
the Lagrangian with the source term is rewritten as
$$
\eqalign{
    {\cal L}_{\rm DGL}&-J\sum_{\alpha =1}^3 |\chi _\alpha |^2
 = {\cal L}_{\rm cl}(\bar \chi ) - 3 J\bar \chi ^2
 - 2 \bar \chi  [ 2\lambda  (\bar \chi ^2-v^2) + J ]
  \sum_{\alpha =1}^3  \tilde \chi _\alpha  \cr
 &- {1 \over 4} (\partial_{\mu}\vec{B}_{\nu}
            - \partial_{\nu}\vec{B}_{\mu} )^2
            + {1 \over 2} m_B^2 \vec{B}_{\mu}^2
 +  \sum_{\alpha=1}^3 [ (\partial_{\mu} \tilde{\chi}_\alpha)^2
      - m_{\chi}^2  \tilde{\chi}_{\alpha}^2 ] \cr
 &+ \sum_{\alpha=1}^3  \{
    g^2(\vec \epsilon_\alpha  \cdot \vec B_\mu )^2
    (\tilde \chi _\alpha ^2+2\bar \chi  \tilde \chi _\alpha )
   - \lambda (4 \bar \chi \tilde{\chi}_{\alpha}^3 + \tilde{\chi}
   _{\alpha}^4) \},
}
\eqn\Lb
$$
where ${\cal L}_{\rm cl}(\bar \chi )$ is the classical part,
$$
{\cal L}_{\rm cl}(\bar \chi ) = -3\lambda (\bar \chi ^2-v^2)^2.
\eqn\Lcl
$$
Here, the masses of $\tilde \chi _\alpha $ and $\vec B_\mu $ are given by
$$
   m_\chi ^2 = 2\lambda (3 \bar \chi ^2-v^2) + J = 4\lambda \bar \chi ^2,
   \hbox{\quad} m_B^2 = 3 g^2 \bar \chi ^2,
\eqn\Ma
$$
where we have used the relation between the mean field
$\bar \chi$ and the source $J$,
$$
J= -2\lambda (\bar \chi ^2-v^2).
\eqn\JCHI
$$
This relation is obtained by the condition that the linear
term of $\tilde{\chi}_{\alpha}$ vanishes.
It is remarkable that the scalar mass $m_\chi $ is always
real owing to the source $J$ even in the negative-curvature
region of the classical potential, $\bar \chi  < v / \sqrt{3}$.

Integrating over $\vec B_\mu $ and $\tilde \chi _\alpha $
by neglecting the higher order terms of the fluctuations,
we obtain the partition functional,
$$
Z[J] = \exp{\left( i \int d^4x
       \{{\cal L}_{\rm cl}(\bar \chi ) - 3J\bar \chi^2 \} \right) }
       [\hbox{Det} (iD_B^{-1})]^{-1}
       [\hbox{Det} (iD_{\chi}^{-1})]^{-3/2},
\eqn\Za
$$
where the exponents, $-1$ and $-3/2$, originate from the
numbers of the internal degrees of freedom.
Here, $D_B$ and $D_\chi $ are the propagators
of $\vec B_\mu $ and $\tilde \chi _\alpha $ in the QCD-monopole
condensed vacuum,
$$
D_B =\left(g_{\mu \nu }-{ k_\mu k_\nu  \over m_B^2 } \right)
{i \over k^2- m_B^2 +i\epsilon }, \hbox{\quad}
D_\chi ={-i \over k^2-m_\chi ^2+i\epsilon }
\eqn\PRa
$$
in the momentum representation.
Hence, the effective action is given by the Legendre transformation
[\cheng],
$$
\Gamma (\bar \chi ) = -i \ln Z[J] + \int d^4x 3 J \bar \chi^2
               =  \int d^4x {\cal L}_{\rm cl}(\bar \chi )
                     +i \ln {\rm Det} (i D_B^{-1})
                     + {3 \over 2} i \ln {\rm Det} (i D_\chi ^{-1}).
\eqn\Ga
$$
The functional determinants are easily calculable in the momentum space,
and we obtain the formal expression of the effective potential [\cheng],
$$
\eqalign{
 V_{\rm eff}(\bar \chi) =
 {-\Gamma (\bar \chi ) / \int d^4x}
    = 3 \lambda (\bar \chi^2 - v^2 )^2
    &+ 3 \int {d^4k \over i(2\pi )^4 }
    \ln(m_B^2-k^2-i\epsilon ) \cr
    &+ {3 \over 2} \int {d^4k \over i(2\pi )^4 }
    \ln(m_\chi ^2-k^2-i\epsilon ).
}
\eqn\Vv
$$

In the finite-temperature system [\kapusta],
the partition functional $Z$ is described by
the Euclidean variables; $x_{0}=-i\tau$, and the
upper bound of the $\tau$ integration is $\beta=1/T$
with $T$ being the temperature.
Then, the $k_0$-integration in the functional determinant
becomes the infinite sum over the Matsubara frequency [\kapusta].
The effective potential at finite temperatures
physically corresponds to the thermodynamical potential,
and is given by
$$
\eqalign{
 V_{\rm eff}(\bar \chi;T) = 3 \lambda (\bar \chi^2 - v^2 )^2
    &+ 3 T \sum_{n=-\infty }^\infty  \int {d^3k \over (2\pi)^3 }
    \ln\{(2n\pi T)^2 + k^2 + m_B^2\} \cr
    &+ {3 \over 2} T \sum_{n=-\infty }^\infty  \int {d^3k \over (2\pi)^3 }
    \ln\{(2n\pi T)^2 + k^2 + m_\chi ^2\}
}
\eqn\Va
$$
in the DGL theory.
Performing the summation over $n$ and the angular integration,
we obtain the final expression of the effective potential at
finite temperatures,
$$
\eqalign{
V_{\rm eff}(\bar \chi ;T) =   3 \lambda ( \bar \chi^2 - v^2 )^2
          &+ 3 {T \over \pi^2} \int_0^\infty  dk k^2 \ln{
           \left(  1 - e^{ - \sqrt{ k^2 + m_B^2}/T }
           \right)  } \cr
          &+ {3 \over 2} {T \over \pi^2}\int_0^\infty  dk k^2 \ln{
           \left(  1 - e^{ - \sqrt{ k^2 + m_{\chi}^2}/T }
           \right)  },
}
\eqn\Vb
$$
where $m_B$ and $m_\chi $ are functions of $\bar \chi $
as shown in Eq.{\Ma}.
Here, we have dropped the $T$-independent part
(quantum fluctuation), because
we are interested in the
thermal contribution to the QCD vacuum [\monden].

We show in Fig.1 the effective potential at various temperatures
(thermodynamical potential), $V_{\rm eff}(\bar \chi ;T)$,
as a function of the QCD-monopole condensate $\bar \chi $,
an order parameter for the color confinement.
The parameters, $\lambda$ = 25, $v$ = 0.126GeV
and $g$ = 2.3, are extracted by fitting the static potential
in the DGL theory to the Cornell potential [\SuST].
\foot{
We examined several possible parameter sets, and
found a small parameter dependence on our results shown
in this paper.
}
These values provide $m_B$ = 0.5GeV and $m_{\chi}$ = 1.26GeV
at $T$ = 0.
The (local-)minimum point of $V_{\rm eff}(\bar \chi ;T)$
corresponds to the physical (meta-)stable vacuum state.
As the temperature increases,
the broken dual gauge symmetry tends to be restored,
and the QCD-monopole condensate in the physical vacuum,
$\bar \chi _{\rm phys}(T)$, is decreased.
A first order phase transition is found at the thermodynamical
critical temperature, $T_C \simeq 0.49$ GeV,
and the QCD vacuum becomes trivial, $\bar \chi _{\rm phys}(T)=0$,
for $T \ge T_C$.
This phase transition is regarded as the
deconfinement phase transition,
because there is no confining force among colored particles
in the QCD vacuum with $\bar \chi _{\rm phys}(T)=0$.

We show the behavior of the QCD-monopole condensate
in the physical vacuum, $\bar \chi _{\rm phys}(T)$,
as a function of the temperature $T$ in Fig.2.
One finds $\bar \chi _{\rm phys}$= 0.126 GeV at $T=0$,
and the QCD-monopole condensate decreases monotonously up to
$\bar \chi _{\rm phys}$ = 0.07 GeV at
the upper critical temperature $T_{\rm up} = $ 0.51 GeV,
where the minimum at finite $\bar \chi $
disappears in $V_{\rm eff}(\bar \chi ;T)$.
On the other hand, the local minimum is developed at $\bar \chi $ = 0
in $V_{\rm eff}(\bar \chi ;T)$ above the lower
critical temperature $T_{\rm low} = $ 0.38 GeV,
which is analytically obtained by
using the high-temperature expansion [\kapusta,\monden],
$$
T_{\rm low}=2v\left( { 6\lambda  \over 2\lambda +3g^2} \right)^{1/2}.
\eqn\LCT
$$
The minimum value of $V_{\rm eff}(\bar \chi ;T)$
at $\bar \chi =0$ becomes deeper than that
at finite $\bar \chi $ above the thermodynamical critical temperature
$T_C$ = 0.49 GeV.
Here, we get the first-order phase transition because we have
considered full orders in $\bar \chi ^2$ as shown in Eq.{\Vb}.
On the other hand, Monden et al. [\monden]
did not get the first-order phase transition due to the
use of only the lowest order in $\bar \chi ^2$
in the high-temperature expansion [\kapusta], and therefore
they had to introduce the cubic term in $\chi _\alpha $ in
the Lagrangian.

Here, we consider the possibility of the temperature dependence
on the parameters ($\lambda $,$v$) in the DGL theory.
The critical temperature, $T_C$ = 0.49 GeV, seems much larger
than the one of the lattice QCD prediction,
$T_C \simeq 0.2$GeV [\rothe].
However, we should remember that the self-interaction term of $\chi _\alpha $
has been introduced phenomenologically in the DGL Lagrangian.
In particular, the asymptotic freedom behavior of QCD leads to a
possible reduction of the self-interaction among QCD monopoles
at high temperatures.
Hence, we use a simple ansatz for the temperature dependence on $\lambda $,
$$
  \lambda (T) = \lambda  \left( {T_C - aT \over T_C}  \right),
\eqn\Lam
$$
keeping the other parameter $v$ constant.
Here, the constant $a$ is determined as $a=0.96$
so as to reproduce $T_C=0.2$GeV. (We take $\lambda (T)=0$ for $T>T_C/a$.)
The results for the monopole condensate
$\bar \chi _{\rm phys}(T)$ are shown in Fig.3.
The qualitative behavior is the same as in the
above argument with a constant $\lambda $.
We find a weak first-order phase transition in this case also.
Here, we find a large reduction of the
self-interaction of the QCD monopoles near the critical
temperature $T_C$: $\lambda (T \simeq T_C) \simeq 1$
is considerably smaller than $\lambda (T=0)=25$.
It would be an interesting subject to examine
such a large reduction of $\lambda (T)$ near $T_C$
from the study of the monopole action in the lattice QCD [\latt].

Next, we investigate the variation of the masses of the
dual gauge field $\vec B_\mu $ and the QCD-monopole field
$\tilde \chi _\alpha $ at finite temperatures.
Here, $\vec B_\mu $ and $\tilde \chi _\alpha $ would appear as
the color-singlet glueball field with $1^+$ and $0^+$, respectively
[\suzuki,\Kro,\TSS].
The glueball masses, $m_B$ and $m_\chi $,
at the finite temperature $T$ are given by
$$
   m_B(T) = \sqrt{3} g \bar \chi _{\rm phys}(T),
   \hbox{\quad}
   m_\chi (T) = 2\sqrt{\lambda (T)}\bar \chi _{\rm phys}(T)
\eqn\GBM
$$
as shown in Eq.{\Ma}.
In Fig.4, We show $m_B(T)$ and $m_\chi (T)$ as functions of the
temperature $T$ using variable $\lambda (T)$ in Eq.{\Lam}.
(Their behaviors are almost the same as the case of a
constant $\lambda $ except for the difference of the value of $T_C$.)
It is worth mentioning that $m_B(T)$ and $m_\chi (T)$
drop down to $m_B, m_\chi  \sim T_C$(=0.2GeV)
from $m_B, m_\chi  \sim$ 1 GeV near the critical temperature $T_C$.
In other words,
the QCD phase transition occurs at the temperature satisfying
$m_B, m_\chi  \simeq T$,
which seems quite natural because the thermodynamical factor
$
1 /\{ \exp(\sqrt{k^2+m^2}/T) \pm 1 \}
$
becomes relevant only for $m \lsim T$.
Thus, our result predicts a large reduction
of the glueball masses, $m_B$ and $m_\chi $, near the
critical temperature $T_C$.
It is desirable to study the change of the glueball masses
at finite temperatures, especially near $T_C$,
in the lattice QCD simulation with the larger lattice size and
the higher accuracy.

We investigate the string tension $k$ at finite temperatures,
since $k$ is one of the most important variables for
the color confinement,
and controls the hadron properties through the inter-quark potential.
We use the expression of the string tension $k(T)$
provided by SST [\SuST],
$$
k(T) = {e^2 m_B^2(T) \over 24 \pi}
\ln \left(  {m_B^2(T) + m_\chi^2(T) \over m_B^2(T)} \right),
\eqn\STRa
$$
where the glueball masses $m_B(T)$ and $m_\chi(T)$
are given by Eq.{\GBM}.
The results are shown in Fig.5 as a function
of the temperature $T$.
In the case of constant $\lambda $, the string tension $k(T)$
decreases very gradually up to the temperature,
$T_{\rm up}$= 0.51 GeV.
On the other hand, in the case of variable
$\lambda (T)$, the string tension $k(T)$ decreases rapidly with
temperature,
and $k(T)$ drops down to zero around $T_C = $ 0.2 GeV.
Hence, one expects a rapid change of
the masses and the sizes of the quarkonia
according to the large reduction of $k(T)$
at high temperatures.
We plot also the results of the lattice QCD simulation
in the pure gauge [\gao] by black dots
near and below the critical temperature, $T_C = $ 0.2 GeV.
We find our results with variable $\lambda (T)$ agree with the lattice
QCD data.

We discuss further the temperature dependence
of the parameters ($\lambda $,$v$) in the DGL theory.
Definitely, we should follow the lattice QCD data
for this determination as the case of the Ginzburg-Landau
theory of superconductors extracting the temperature dependence
from experiments.
Since there exists the lattice QCD data on the string tension
$k$ [\gao], we try to reproduce $k$ by taking a simple ansatz on
$\lambda $ and $v$.
We try the following ansatz,
$$
  B(T) \equiv 3\lambda (T)v^4(T)
  = 3\lambda v^4 \left( {T_C - aT \over T_C}  \right),
\eqn\LamZ
$$
where the constant $a$ is determined so as to reproduce $T_C=0.2$GeV.
(We take $B(T)=0$ for $T>T_C/a$.)
The variable $B(T)$ corresponds to the bag constant, the energy-density
difference between the nonperturbative vacuum ($|\chi _\alpha | \ne 0$)
and the perturbative vacuum ($|\chi _\alpha |=0$) in the DGL theory;
see Eq.{\Vb}.
The ansatz {\LamZ} suggests the reduction of the bag constant
at high temperatures, which provides the swelling of hot hadrons
by way of the bag-model picture.
Since we have already examined a typical case for variable
$\lambda (T)$ keeping $v$ constant, we show here another typical case
for variable $v(T)$ keeping $\lambda $ constant.
The string tension $k(T)$ in the variable $v(T)$ case with $a=0.97$
is shown by the dashed line in Fig.5.
We find almost an identical result and find again a good agreement
with the lattice QCD data.
Other combinations on $\lambda (T)$ and $v(T)$ under the relation {\LamZ}
also provide equally good results on $k(T)$.

Finally, we investigate the relation between the scalar glueball mass
$m_\chi (T)$ and the string tension $k(T)$.
For variable $\lambda (T)$ keeping $v$ constant,
one finds from Eq.{\STRa} an approximate relation,
$$
{m_\chi (T) \over \sqrt{k(T)}} \simeq {(24\pi )^{1/2} \over e}
\simeq 1.6,
\eqn\GBMx
$$
near the critical temperature $T_C$.
On the other hand, for variable $v(T)$ keeping $\lambda $ constant,
the glueball masses at finite temperatures,
$m_B(T)$ and $m_\chi (T)$, are shown by the dashed line in Fig.4,
and Eq.{\STRa} leads to a simple relation,
$$
{m_\chi (T) \over \sqrt{k(T)}}
= \left( {2\lambda  \over \pi  \ln \{ (3g^2+4\lambda )/3g^2 \}} \right)^{1/2}
\simeq 3.0,
\eqn\GBMy
$$
for the whole region of $T$.
Thus, the DGL theory suggests a proportional relation between
the scalar glueball mass and the square root of the string
tension at least near $T_C$.
It is worth mentioning that Engels et al. [\karsch]
obtained a similar relation, $m_{\rm GB}(T)= (1.7 \pm 0.5) \sqrt{k(T)}$,
for the lowest scalar glueball at finite temperatures
from the thermodynamical study on the SU(2) lattice gauge theory.
Eqs.{\GBMx} and {\GBMy} can be examined from the
thermodynamical study on the glueball mass in the lattice QCD,
which may also reveal $T$-dependence on the parameters in the
DGL theory.

We have studied the properties of the pure-gauge QCD vacuum
at finite temperatures in the dual Ginzburg-Landau (DGL) theory, where
the color confinement is realized through the dual Higgs mechanism.
We have formulated the effective potential at finite temperatures
(thermodynamical potential)
using the path-integral formalism.
We have used the quadratic source term instead of the
linear source term.
The use of the quadratic source term overcomes the problem of the
imaginary scalar mass, which is encountered in the case of
the linear source term as is well-known in the $\phi^4$ theory.

We have found the reduction of the QCD-monopole
condensate at finite temperatures,
and have found a first-order deconfinement phase transition
at the critical temperature $T_C \simeq $ 0.49GeV
using the temperature-independent parameters.
The QCD-monopole condensate vanishes and
the broken dual gauge symmetry is restored above $T_C$.
We have considered the temperature dependence of the QCD-monopole
self-interaction noting $T_C=0.2$GeV
as the lattice QCD simulation indicates.
We have found a large reduction of the
QCD-monopole self-interaction near the critical temperature.
We have investigated the temperature dependence of the
glueball masses, $m_B$ and $m_\chi $,
and have found their large reduction
near the critical temperature $T_C$: $m_B, m_\chi  \sim T_C$.
We have calculated also the string tension at finite temperatures.
The results agree with the lattice QCD data both in
the variable $\lambda (T)$ and in the variable $v(T)$ cases.

In particular, the glueball mass reduction at high temperatures
would be an important ingredient in the QCD phase transition.
In the pure gauge, there are only glueball excitations
with the large masses ($\gsim$ 1GeV) at low temperatures
[\rothe,\karsch],
and therefore it seems unnatural that the QCD phase transition
takes place at a small critical temperature, $T_C \simeq$ 0.2GeV.
This problem would be explained by
the large reduction of the glueball mass near the
critical temperature as is demonstrated in this paper.
In other words, this glueball-mass reduction may
determine the magnitude of the critical temperature $T_C$
in the QCD phase transition.

It is also very interesting
to investigate the thermal effect of the dynamical quark
in the DGL theory, especially in terms of the
chiral phase transition [\rothe,\kapusta].
The chiral-symmetry restoration as well as the deconfinement
phase transition is expected to happen in the DGL theory,
because our previous works [\SuST,\TSS,\SaST] showed the
close relation between the color confinement and the dynamical
chiral-symmetry breaking in the DGL theory.
In the presence of dynamical quarks,
there would be two important ingredients.
One is the glueballs, $\vec B_\mu $ and $\chi _\alpha $,
characterizing the confinement properties of the QCD vacuum.
Their mass would decrease near $T_C$ according to the
deconfinement phase transition.
The other is the light hadrons as pions,
which have a close relation to
the spontaneous chiral-symmetry breaking [\cheng].
The thermal contribution of light hadrons as pions would lower
the critical temperature $T_C$.
The lattice QCD simulations showed
the close values of the critical temperatures
between the pure-gauge case ($T_C\simeq$ 0.2GeV)
and the case with dynamical quarks ($T_C \simeq$ 0.15GeV) [\rothe].
These close values of $T_C$ would be explained
if the glueball-mass reduction at high temperatures
plays the dominant role in the QCD phase transition
even in the presence of dynamical quarks.
Such a conjecture on the glueball mass reduction
would provide an interesting subject in the lattice QCD
[\karsch].

We would like to thank H.~Monden and S.~Sasaki for stimulating
discussions on the DGL theory.
One of the authors (H.S.) is supported by the Special Researchers'
Basic Science Program at RIKEN.

\refout

\figout

\end